\documentclass[12pt]{iopart}
\usepackage{iopams} 
\usepackage{graphicx}
\begin{document}

\title[]{Suspended silicon nitride thin films with enhanced and electrically tunable reflectivity}

\author{Bhagya Nair$^1$, Andreas Naesby$^1$, Bjarke R. Jeppesen$^2$ and Aur\'{e}lien Dantan$^1$}

\address{$^1$Department of Physics and Astronomy, University of Aarhus, DK-8000 Aarhus C, Denmark}
\address{$^2$Interdisciplinary Nanoscience Center (iNANO), Aarhus University, Gustav Wieds Vej 14, DK-8000, Aarhus C, Denmark}
\ead{dantan@phys.au.dk}

\begin{abstract}
We report on the realization of silicon nitride membranes with enhanced and electrically tunable reflectivity. A subwavelength one-dimensional grating is directly patterned on a suspended 200 nm-thick, high stress commercial film using electron beam lithography. A Fano resonance is observed in the transmission spectrum of TM polarized light impinging on the membrane at normal incidence, leading to an increase in its reflectivity from 10\% to 78\% at 937 nm. The observed spectrum is compared to the results of rigorous coupled wave analysis simulations based on measurements of the grating transverse profile through localized cuts of the suspended film with a Focused Ion Beam. By mounting the membrane chip on a ring piezoelectric transducer and applying a compressive force to the substrate we subsequently observe a shift of the transmission spectrum by 0.23 nm.
\end{abstract}

%
%
%
%
%

\section{Introduction}

Suspended thin films are widely used in photonics and sensing applications. Among them, silicon nitride films deposited with low pressure chemical vapor deposition benefit from excellent optical properties, high refractive index and ultralow loss in the visible and near-infrared range, as well outstanding mechanical properties, high frequency/quality factor mechanical resonances and high burst pressure. Due to these features as well as their low mass, thin SiN films with thickness ranging from tens to a few hundreds of nanometers have been applied in a number of cavity optomechanics investigations~\cite{Thompson2008,Jayich2008,Wilson2009,Camerer2011,Karuza2012,Purdy2013,Sawadsky2015,Xu2016,Nielsen2017}. Increasing the otherwise relatively low reflectivity of such membranes without increasing their mass is beneficial for cavity optomechanics with single~\cite{Kemiktarak2012apl,Kemiktarak2012njp,Bui2012,Stambaugh2015,Bernard2016,Norte2016,Reinhardt2016,Chen2017,Moura2018} or multiple resonators~\cite{Bhattacharya2008,Hartmann2008,Xuereb2012,Xuereb2014,Li2016,Nair2016,Piergentili2018}, as well as for sensing~\cite{Bruckner2010,Guo2017,Naesby2018} or lasing~\cite{Huang2008,Kemiktarak2014,Yang2015} applications. Such an increase in reflectivity for films with subwavelength thickness can be realized by patterning the film with a suitable subwavelength periodic structure, such as a high contrast grating~\cite{ChangHasnain2012} or a photonic crystal structure~\cite{Zhou2014}. The interference between the incoming light and guided modes in the structure allows for tailoring the optical properties of the film. In particular, the destructive interference occuring for specific wavelengths/polarizations can give rise to the appearance of either broad- or narrowband Fano-like resonances in the reflectivity spectrum~\cite{Miroshnichenko2010}. In a broader context, resonant waveguide gratings and photonic crystal structures are relevant for a variety of optical applications, including waveguide coupling, optical filtering, polarizers, spectrometry, biosensing, solar cells, photodetection, lasing, etc.~\cite{Quaranta2018}.

Enhanced reflectivity of suspended silicon nitride membranes has been observed both for two-dimensional photonic crystals~\cite{Bui2012,Bernard2016,Norte2016,Chen2017,Moura2018,Gartner2018} and for one-dimensional subwavelength gratings (SG)~\cite{Kemiktarak2012apl,Kemiktarak2012njp,Stambaugh2015} patterning. We focus here on the realization of polarization-dependent, one-dimensional SGs similar to those of Refs.~\cite{Kemiktarak2012apl,Kemiktarak2012njp,Wang2015,Stambaugh2015}, albeit by following the approach of~\cite{Chen2017}~and \textit{directly} patterning a $50\times 50$ $\mu$m$^2$ subwavelength grating on a commercial, 200 nm-thick silicon nitride membrane. A Fano resonance is observed  in the transmission spectrum of transverse magnetic (TM) polarized light impinging at normal incidence, showing an increase in reflectivity at 937 nm from 10\% for an unpatterned membrane to 78\% for a patterned one.  By using localized cuts of the structure with a Focused Ion Beam~\cite{Ierardi2014} we also show that the grating transverse grating profile of the suspended film can be measured. We compare the observed transmission spectrum with the results of Rigorous Coupled Wave Analysis numerical simulations taking into account the grating profile and discuss collimation broadening and finite grating size effects. 

Last, by mounting the membrane chip on a ring piezoelectric transducer and applying a compressive force to the corners of the substrate we show as a proof of principle that it is possible to shift the transmission spectrum by 0.23 nm towards lower wavelengths. Optimizing and increasing the piezoelectric compressive force further would make such patterned membranes interesting for realizing tunable optical filters~\cite{Wang2015}, waveplates~\cite{Mutlu2012} or strongly focusing lenses~\cite{Fattal2010,Lu2010,Klemm2013}. Increasing the patterned area~\cite{Chen2017,Moura2018} and the reflectivity would also make them attractive for vertical-cavity surface-emitting lasers~\cite{Huang2008,Zhou2008} or optical sensors~\cite{Guo2017,Naesby2018} for biophysics~\cite{Schuler2009,Dong2017} and biomedical~\cite{Leinders2015,Sinha2017} applications. Last, combined with piezoelectric actuation of their mechanical modes~\cite{Wu2018,Naserbakht2019}, such enhanced and electrically tunable reflectivity membranes would be particularly interesting for improving and tailoring the optomechanical response of arrays of nanomembranes~\cite{Nair2017,Piergentili2018,Gartner2018} and for investigating collective and strong coupling optomechanics~\cite{Xuereb2012,Xuereb2013,Xuereb2014,Xuereb2015,Cernotik2019}.

\section{Fabrication}

\begin{figure}[h!]
\includegraphics[width=\columnwidth]{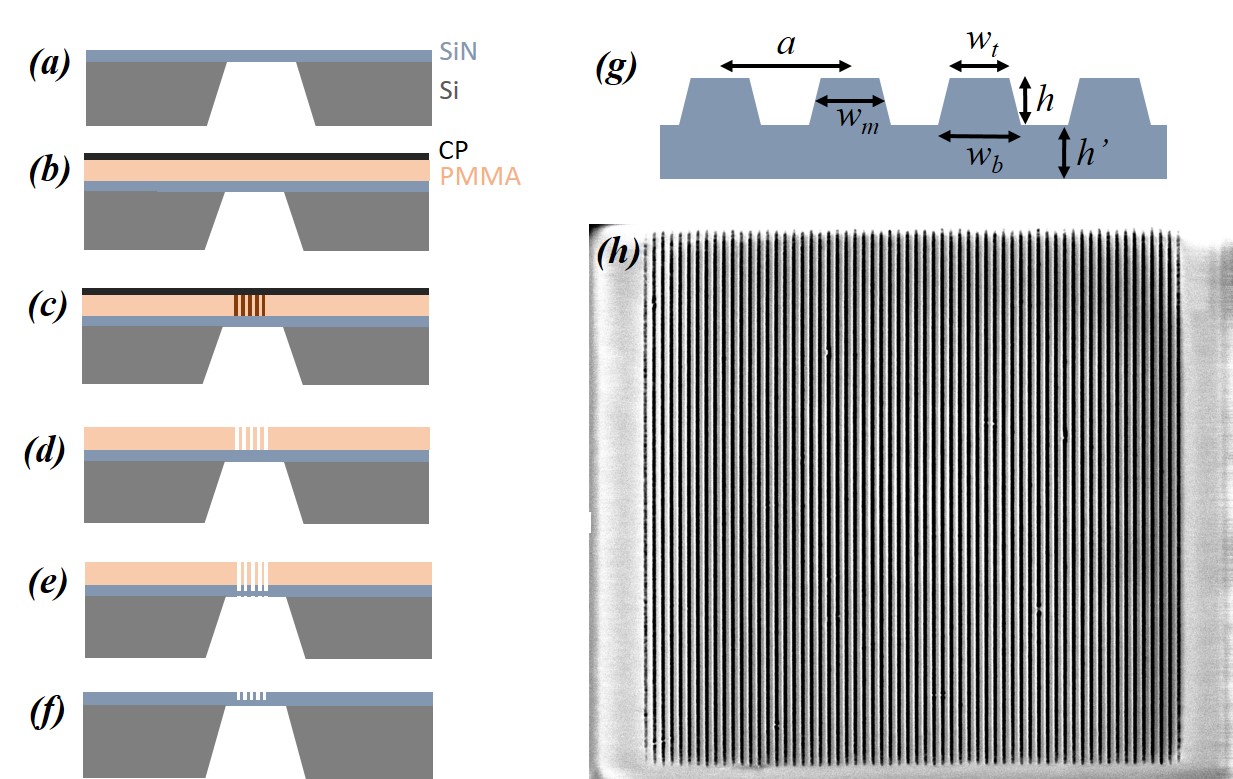}
\caption{(a-f) SG fabrication process. (a) Suspended SiN membrane on Si, (b) Coating with PMMA and CP, (c)  EBL,  (d) Development, (e) Etching, (f) PMMA removal. (g) Schematic cross section of a SG with tilted walls. (h) Topview SEM image of the $50\times 50$ $\mu$m$^2$ SG patterned area.}
\label{fig:fab}
\end{figure}

The stochiometric silicon nitride membranes used in this work are commercial~\cite{Norcada}, high stress ($\sim$GPa), 0.5~mm-square and 200~nm-thick films on a 5~mm-square, 500~$\mu$m-thick silicon frame. The steps for patterning the SG are depicted in Fig.~\ref{fig:fab}a and are as follows: after cleaning with an O$_2$ plasma the sample is spin-coated with a 9\% 950k molecular weight PMMA resist and a conductive polymer layer (SX-AR-PC 5000/90.2) to avoid charging effects during Electron Beam Lithography (EBL). A $50\times50$ $\mu$m$^2$ grating mask with a target period and finger width is then written by EBL at 30 kV. After writing the conductive polymer layer is removed by immersion in deionized water and the PMMA resist is developed in a solution of 3:7 H$_2$O:IPA (Iso-propyl alcohol). The membrane is then etched in a STS Pegasus ICP DRIE system using reactive ion etching with C$_4$F$_8$ (flow rate 59 sccm) and SF$_6$ (flow rate 36 sccm) for 170 seconds at 800 W. The PMMA layer is removed in acetone followed by rinsing in IPA and N$_2$ blow drying. The sample is finally cleaned by means of the O$_2$ plasma again. 

The etching parameters are observed to be quite critical for the success of the patterning, and etching only a fraction of the 200 nm films is determinant in avoiding rupture/damage during the process. A top-view SEM picture of the resulting grating is shown in Fig.~\ref{fig:fab}h. From such SEM images the period and grating finger width are estimated to be $a=(810\pm 15)$ nm and $w=(450\pm 20)$ nm, respectively, where the uncertainties come from the imaging system calibration. The latter number is potentially deceiving, however, as the vertical profile of the grating is not expected to be rectangular and the observed mean finger width depends on the saturation level of the SEM image.

\begin{figure}[h!]
\centering\includegraphics[width=\columnwidth]{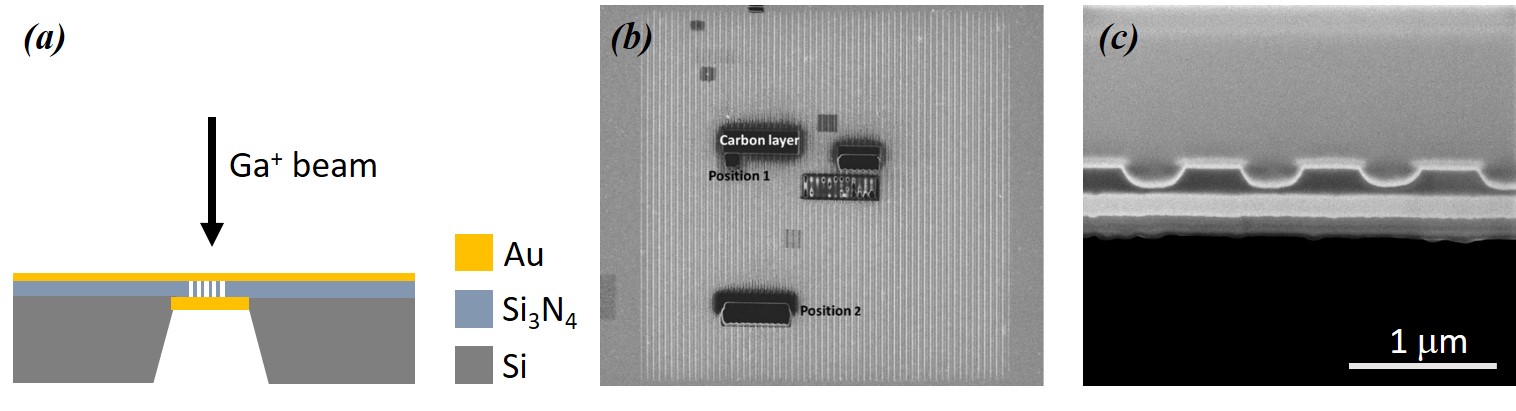}
\caption{(a) Schematic of the FIB cutting setup. (b) SEM topview image of the SG after cutting. (c) SEM cross section view of the SG profile at position 1.}
\label{fig:fib}
\end{figure}

To get more insight into the vertical profile of the grating structure a Focused Ion Beam (FIB) was used to cut through the SG at different locations of the patterned area, as shown in Figs.~\ref{fig:fib}a and b. Before the FIB cut the membrane was made more robust by coating on both the etched and non etched sides with $\sim 50$~nm and $\sim 200$~nm of gold, respectively, and a thick ($\simeq 2\mu$m) carbon layer was deposited at the various cutting positions. The gold layers prevent the structure from fracturing and breaking when exposed to the ion beam. Cuts in several locations of the same grating can then be performed, as illustrated in Fig.~\ref{fig:fib}b. A high energy Ga$^+$ ion beam is focused at the edge of the cutting positions defined by the carbon layers, as shown in Figs.~\ref{fig:fib}a and b. The grating cross section can then be imaged at a 45$^{\circ}$ angle with a SEM microscope (Fig.~\ref{fig:fib}c). As expected, the grating fingers are observed to have smooth edged walls, whose width substantially varies from top to bottom. Analyzing these images yield for this particular sample a period $a= (806\pm 10)$~nm, mean/top/bottom/mean finger widths of $w_m= (512\pm10)$~nm, $w_t= (395\pm14)$~nm and $w_b= (628\pm15)$~nm, respectively. The grating height is $h= (109\pm 8)$~nm and the underlying Si$_3$N$_4$ layer thickness $h'= (87\pm 7)$~nm. The transverse profile obtained from these images can be used as input to the numerical simulations to predict transmission spectra, as will be discussed below.

\section{Optical characterization}

\begin{figure}[h!]
\centering\includegraphics[width=0.8\columnwidth]{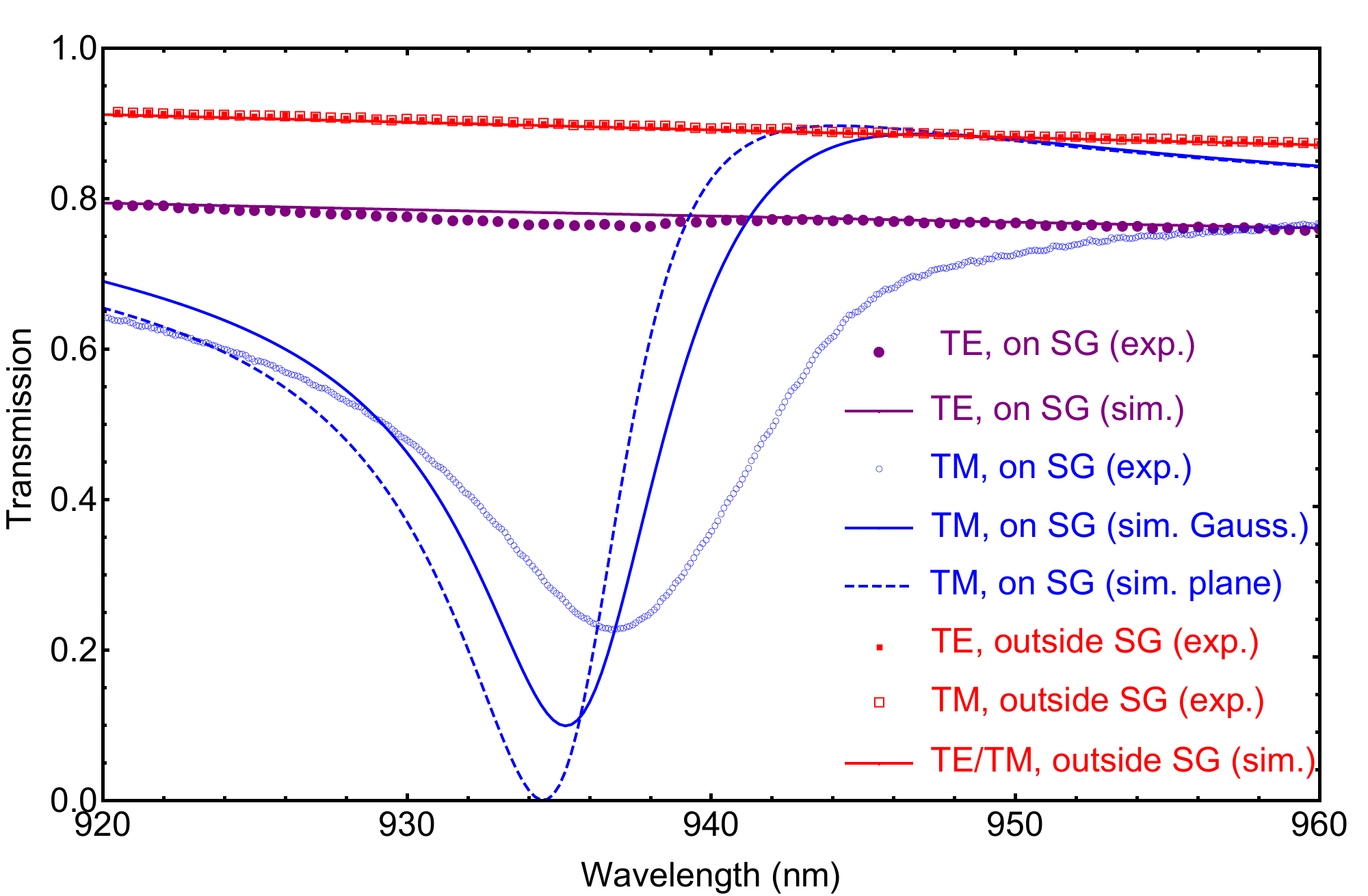}
\caption{Normal incidence normalized transmission of the patterned membrane for TE (full symbols) and TM (open symbols) polarized light incident on (blue and magenta) and outside (red) the patterned area. The solid/dashed lines are the results of the RCWA simulations for the grating parameters given in the text. For TM polarized light incident on the SG the dashed line shows the result of RCWA simulations of a plane wave impinging at normal incidence on an infinite grating, while the solid line shows the result of simulations of a Gaussian beam (superposition of plane waves incident at different incidence angles) on the grating (see text for details).}
\label{fig:trans}
\end{figure}

The transmission of a patterned membrane similar to the one described in the previous section was measured by focusing onto the sample linearly polarized light issued from a tunable external cavity diode laser with a $f=60$ mm focal length achromat doublet (spotsize $\sim$30 $\mu$m). Figure~\ref{fig:trans} shows the normalized transmission measured for TE and TM polarized light impinging at normal incidence either on or outside the patterned area.

Outside the patterned area the transmissivity of the film is independent of the polarization of the light and its level is determined by the film thickness and refractive index. From ellipsometry measurements and broadband transmission spectroscopy~\cite{Nair2017} the thickness and refractive index of unprocessed samples from the same fabrication batch were determined to be $t\simeq 200$ nm and $n\simeq 1.982$ in the plotted range (920-960 nm). Consistently with these values the measured transmissiondecreases from 12 to 8\% in that range. When light impinges on the patterned area an overall small drop in transmission is observed for TE polarized light, on account of the reduced effective thickness of the patterned film. A much more stronger wavelength-dependent drop is observed for TM polarized light around 937 nm, as a result of the interference between the incoming light and guided modes in the structure in this wavelength range. 

To investigate the observed spectrum we performed numerical simulations based on a RCWA approach~\cite{MIST} and using as input parameters the refractive index $n=1.982$ determined previously, the grating period $a=815$ nm and mean finger width $w_m=530$ nm estimated from topview SEM images and the approximate grating profile shown in Fig.~\ref{fig:fab}g with the thicknesses $h=109$ nm and $h'=87$ nm determined from the FIB cut measurements. In this approach a plane wave is incident on an infinite grating, whose unit cell is defined by the previous parameters (20 layers were used to discretize the structure), and the transmitted or reflected fields can be computed. The dashed blue line in Fig.~\ref{fig:trans} shows the result of such a simulation for a plane wave impinging on the grating at normal incidence ($\theta=0$), which predicts a narrower Fano resonance with zero transmission occuring around 934.5 nm. The beam incident on the grating is however a focused Gaussian beam with a waist $w_0\sim 15$ $\mu$m. This beam can be considered as a superposition of plane waves impinging on the grating with various incident angles $\theta$~\cite{Crozier2006,Bernard2016,Moura2018}. The transmittivity of the collimated beam is then computed by performing a weighted average of the RCWA-simulated transmittivities of plane waves incident at different angles, with a Gaussian distribution $e^{-2\theta^2/\theta_D^2}$, where $\theta_D=\lambda/\pi w_0$ is the Gaussian beam divergence angle ($\theta_D\simeq 1.1^{\circ}$ in our case). The resulting spectrum (solid blue line in Fig.~\ref{fig:trans}) shows a shifted and broadened Fano resonance displaying non-zero minimal transmission, as compared to the infinite plane wave case, and is closer to the measured spectrum, although the latter is somehow still more shifted and broadened. Slightly altering the grating parameters in the simulations around the measured values within the systematic uncertainties does not qualitatively change this picture; while collimation broadening can be expected to play a non-negligible role, other effects are likely to contribute to the broadening and shifting of the spectrum. Effects due to the finite size of the grating could also play a role. These could in principle be numerically simulated using full three-dimensional finite element analysis; however, such simulations are computationally heavy and were not performed here. Instead, we note that an estimate of the broadening of a waveguide resonance due to finite size effects can be obtained following analysis of guided-mode resonance filters and couplers~\cite{Brazas1995,Saarinen1995,Boye2000} and is approximately given by $C\lambda a/l$, where $l$ is the grating dimension and $C$ a constant of order unity. In our case ($l=50$ $\mu$m), this broadening would be of the order of 15 nm, but may be expected to be overestimated given the relatively high contrast of the grating structure~\cite{Quaranta2018,Ko2018}. It is however reasonable to think that finite size effects contribute to some extent to the broadening of the observed spectrum. Let us point out that both collimation broadening and finite size effects could be reduced by increasing the size of the patterned area in the future~\cite{Chen2017,Moura2018}. Another possible broadening effect would be inhomogeneities in the grating structure; these could in principle be investigated by applying the method demonstrated in the previous section and performing cuts in several locations of the grating.

\section{Piezoelectric tuning}

\begin{figure}[h]
\centering\includegraphics[width=0.33\columnwidth]{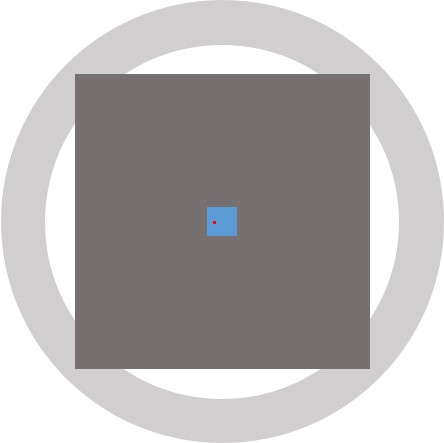}\\
\centering\includegraphics[width=0.65\columnwidth]{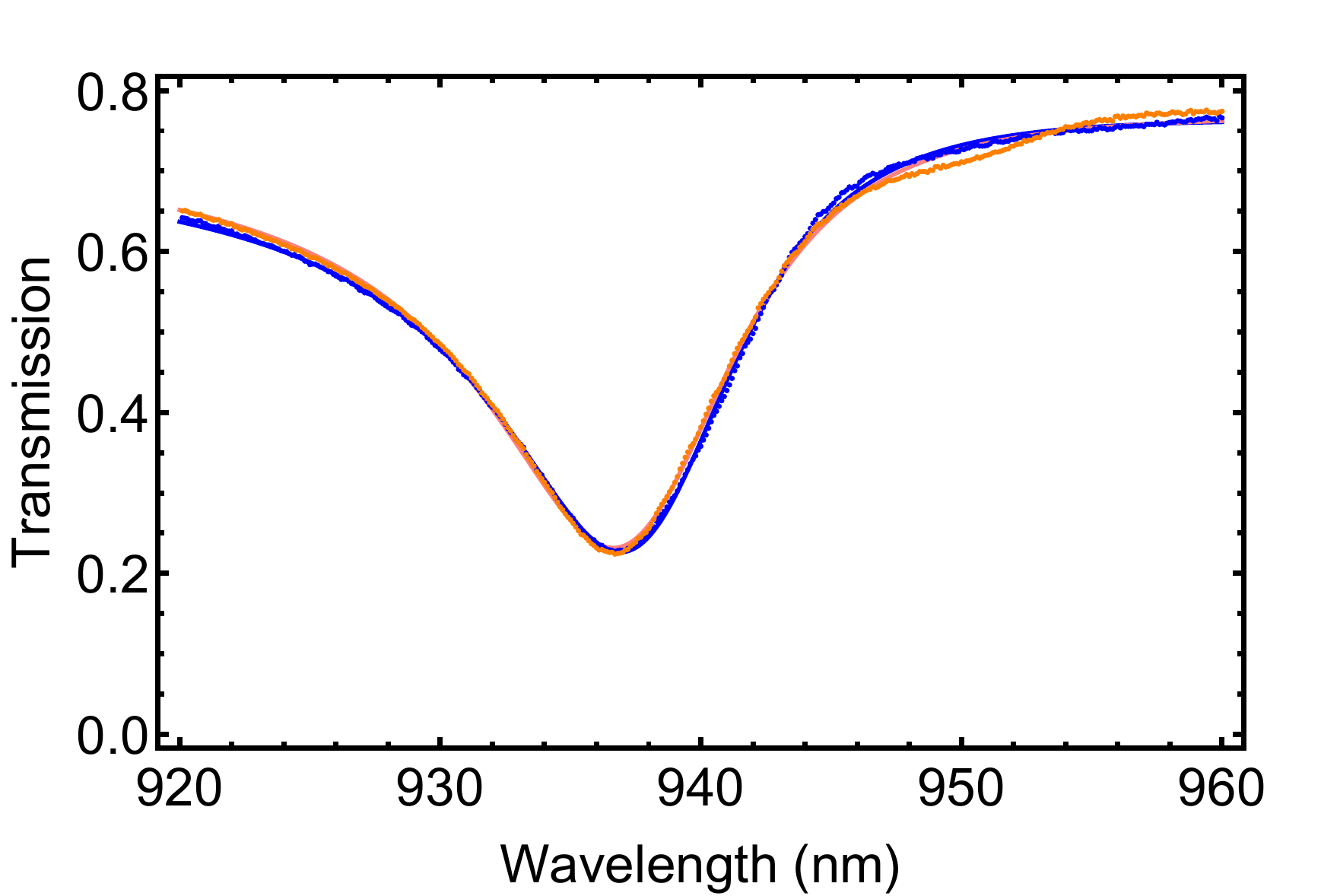}\\
\centering\includegraphics[width=0.65\columnwidth]{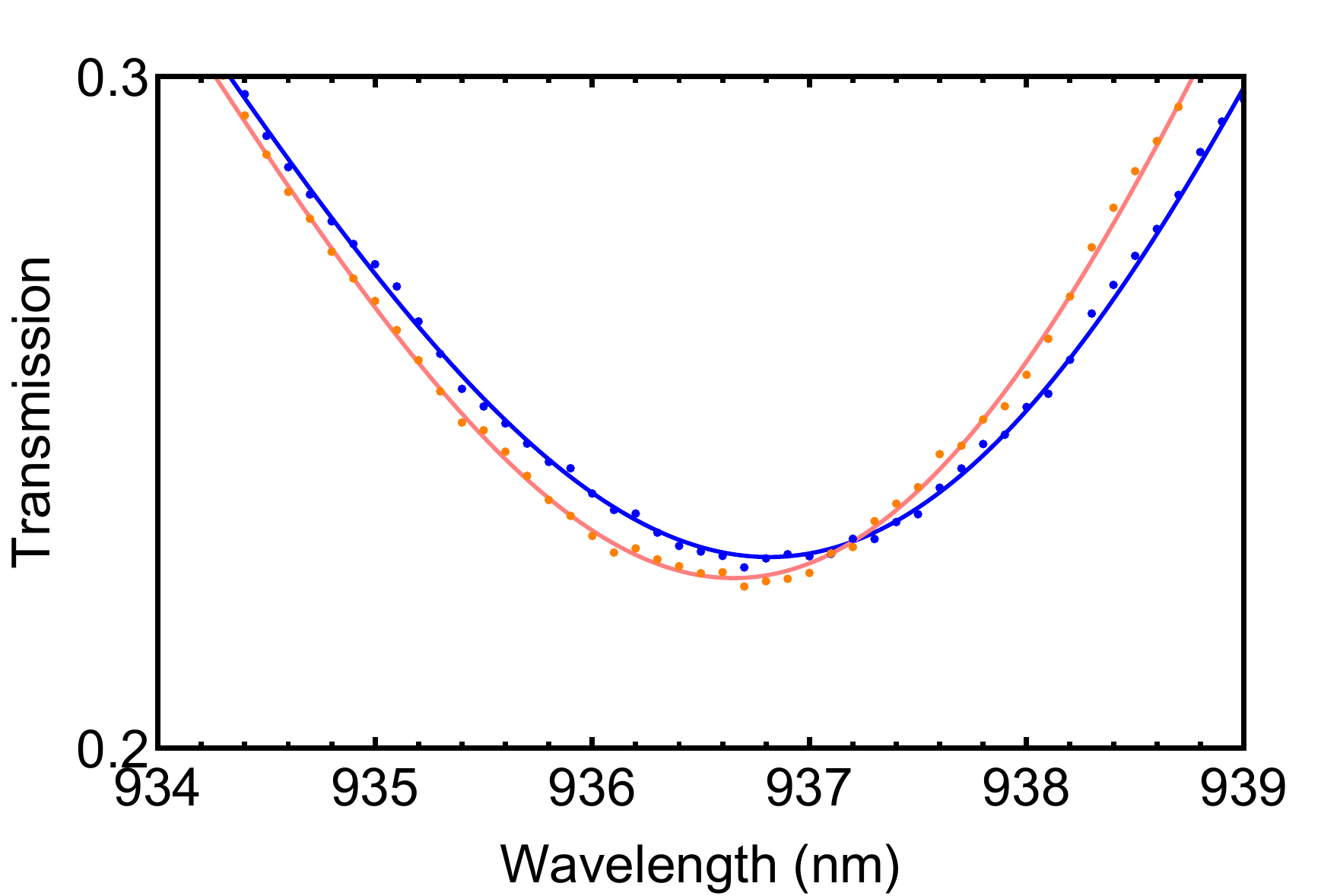}
\caption{Top: Topview schematic (to scale) of the patterned membrane chip mounted on the ring piezoactuator (blue: SiN membrane, dark grey: Si chip, light grey: piezoactuator). The red dot indicate the position of the SG on the membrane. Middle: Normalized transmission spectra for TM polarized light at normal incidence on the SG, with 0 V (blue) and with 180 V (orange) applied to the piezoactuator. The dots are the average of 3 spectra and the statistical errors are smaller than the dot size. The solid lines are the results of fits with a Fano profile. Bottom: zoom of these spectra around the Fano resonance.}
\label{fig:actuation}
\end{figure}

In subsequent experiments three corners of the patterned membrane were glued on a ring piezoelectric actuator (Noliac NAC2123) with 6 mm inner diameter and a specified load-free inner radius contraction of $\Delta r\sim -9$ $\mu$m for an applied 180 V voltage. The grating being offset from the membrane center by about $d\sim160$ $\mu$m its period is slightly reduced by the inward contraction of the SiN film and a global shift of the Fano resonance can then be expected. The same transmission measurements as in the previous section were performed with and without voltage applied to the actuator. The results are shown in Fig.~\ref{fig:actuation} for TM polarized light at normal incidence. The spectrum is clearly shifted towards lower wavelengths when a 180 V voltage is applied. The solid lines show the result of fits with a Fano profile of the form
\begin{equation}
T(\lambda)=A+B\frac{\left(\lambda-\lambda_0+q\gamma\right)^2}{(\lambda-\lambda_0)^2+\gamma^2}
\end{equation}
and match the experimental spectra fairly well. A shift of the resonance frequency $\lambda_0$ of $-0.23\pm 0.03$ nm can be inferred from the fits to the Fano profiles. This is consistent with what can be expected from the resulting reduction in grating period taking into account its off-centered position. The RCWA simulations show that the relative shift in resonance wavelength scales like the relative change in the grating period, i.e. $\Delta\lambda_0/\lambda_0\sim\Delta a/a$. Assuming a linear contraction between the edges of the chip and its center, one has that $\Delta a/a\sim \sqrt{2}(\Delta r/L)(d/L)$, where $2L=5$ mm is the chip lateral dimension. One then gets an expected resonant wavelength shift $\Delta \lambda_0\sim-0.3$ nm, in reasonable agreement with the measured shift.

\begin{figure}[h]
\centering\includegraphics[width=0.65\columnwidth]{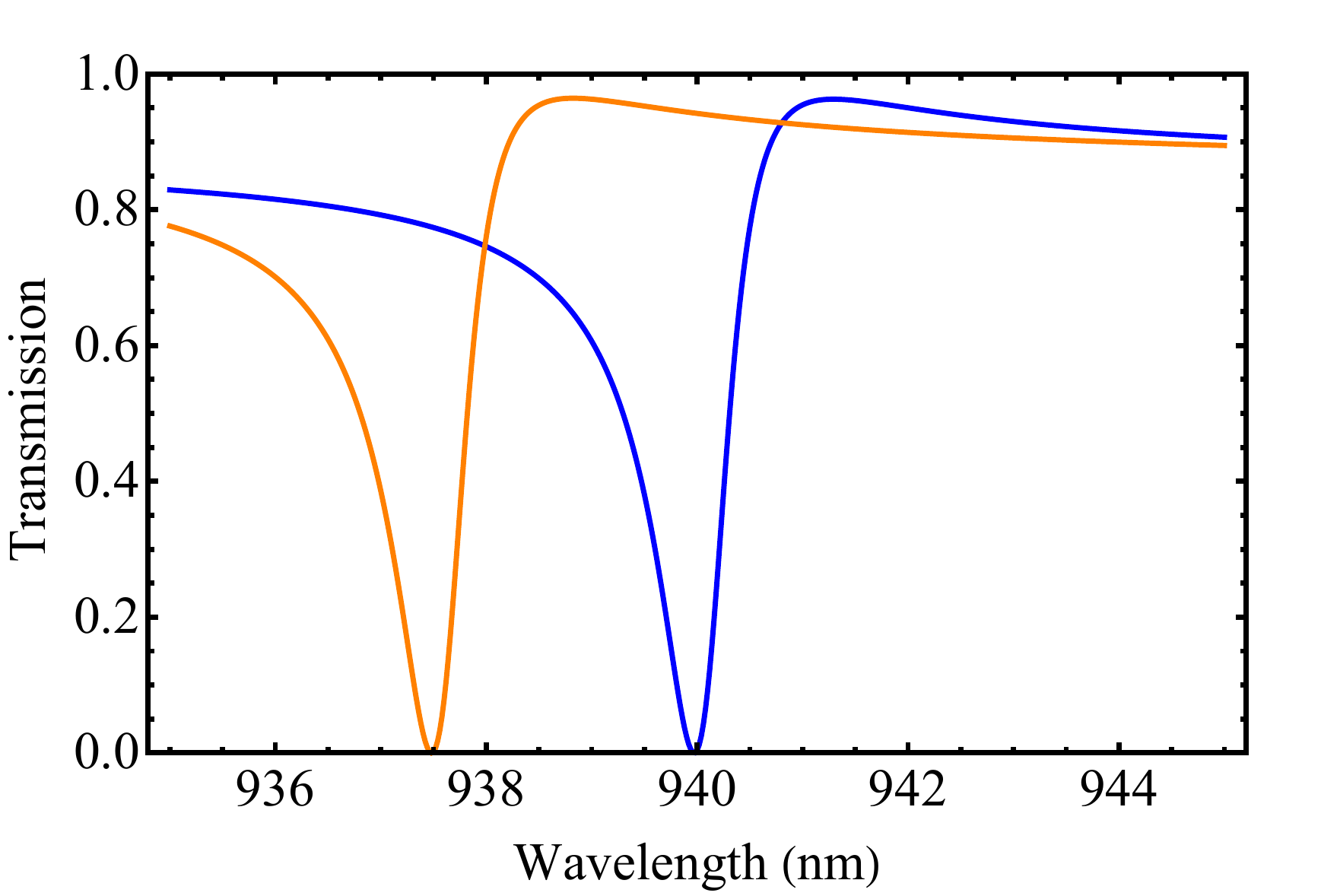}
\caption{RCWA simulated spectra of a high-Q ($\sim1000$) SG (see text for parameters) without (blue) and with (orange) a compression corresponding to a 3 nm grating period change.}
\label{fig:sim_act}
\end{figure}

Let us note that the shift is relatively modest owing to the piezoelectric transducer and geometry used for this device, which was not optimized for this purpose, but rather for optomechanical applications in which preservation of the high quality factor of the mechanical resonances requires minimizing clamping losses as much as possible~\cite{Wu2018,Naserbakht2019}. The 6 mm-inner diameter of the piezoelectric ring actuator was thus chosen only slightly smaller than the 7 mm chip diagonal dimension. For photonics applications, such as low loss tunable optical filters and waveplates, much larger shifts could in principle be obtained by increasing the compression force and optimizing the position of the grating with respect to the piezoelectric transducer. For instance, using larger membrane windows and/or different actuation geometry, as in e.g.~\cite{Wang2015}, could readily yield grating period changes $\Delta a$ of several nanometers. Second, the optical quality factor of the Fano resonance used here is not particularly high ($\sim 90$). Making use of subwavelength gratings with much higher optical quality factor resonances ($Q>14 000$ was reported in~\cite{Zhou2008}) would readily improve the switching contrast of such a filter. For instance, assuming a reasonable ten times larger period change of 3 nm and an optical Q of $\sim 1000$ would give $\Delta\lambda_0/\lambda_0\sim 3\times 10^{-3}\gg 1/Q$ and a switching contrast of about 10 dB. To verify this, the results of RCWA simulations of the transmission of an infinite SG with $a=729$, $w_m=705$ nm, $h=50$ nm and $h'=150$ nm with and without a compression corresponding to such a 3 nm grating period change are shown in Fig.~\ref{fig:sim_act}.

Finally, let us mention that the use of a high optical Q subwavelength grating, even with modest electrical tunability, as the flexible end-mirror of an ultrashort cavity would be interesting for investigating new regimes of cavity optomechanics with strongly-wavelength dependent reflectors~\cite{Naesby2018,Cernotik2019}.

\section{Conclusion}
The possibility to directly pattern one-dimensional subwavelength gratings on commercial, high-stress, 200~nm-thick suspended silicon nitride films was demonstrated. An enhancement of the reflectivity from 10\% to 78\% was observed for 937 nm TM polarized light  focused on a $50\times 50$ $\mu$m$^2$ grating. The broadening and non-zero minimal transmission of the Fano resonance observed in the transmission spectrum were discussed based on RCWA simulations and using transverse profile images of the grating obtained by FIB cutting. Increasing the patterned area and further optimizing of the writing/etching process using the FIB cutting method are expected to reduce collimation broadening and finite grating size effects, and thereby increase the reflectivity. Fine tunability of the optical spectrum by piezoelectric contraction of the suspended film was furthermore evidenced. Such enhanced and electrically tunable reflectivity membranes would be interesting for investigations of collective phenomena in optomechanical arrays of nanomembranes as well as for a number of photonics and optical sensing applications.

\section*{Acknowledgments}
We are grateful to Pia Bomholt Jensen for assistance with FIB cutting and Folmer Lyckegaard for assistance with gold coating, and acknowledge support from the Velux Foundations, the Danish Council for Independent Research (Sapere Aude initiative) and the Carlsberg Foundation.

\bibliographystyle{iopart-num.bst}
\bibliography{biblio_grating_membrane}

\end{document}